# An Approach to Investigate Public Opinion, Views, and Perspectives Towards Exoskeleton Technology

*Nirmalya Thakur[1], Cat Luong[1], Chia Y. Han[1]*

[1] *Department of Electrical Engineering and Computer Science, University of Cincinnati, Cincinnati, OH 45221-0030, U.S.A.*

## ABSTRACT

Over the last decade, exoskeletons have had an extensive impact on different disciplines and application domains such as assisted living, military, healthcare, firefighting, and industries, on account of their diverse and dynamic functionalities to augment human abilities, stamina, potential, and performance in a multitude of ways. In view of this wide-scale applicability and use-cases of exoskeletons, it is crucial to investigate and analyze the public opinion, views, and perspectives towards exoskeletons which would help to interpret the effectiveness of the underlining human-robot, human-machine, and human-technology interactions. The Internet of Everything era of today's living, characterized by people spending more time on the internet than ever before, holds the potential for the investigation of the same by mining and analyzing relevant web behavior, specifically from social media, that can be interpreted to understand public opinion, views, and perspectives towards a topic or set of topics. Therefore, this paper aims to address this research challenge related to exoskeletons by utilizing the potential of web behavior-based Big Data mining in the modern-day Internet of Every-thing era. As Twitter is one of the most popular social media platforms on a global scale – characterized by both the number of users and the amount of time spent by its users on the platform – this work focused on investigating web behavior on Twitter to interpret the public opinion, views, and perspectives towards exoskeleton technology. A total of approximately 20,000 tweets related to exoskeletons were used to evaluate the effectiveness of the proposed approach. The results presented and discussed uphold the efficacy of the proposed approach to interpret and analyze the public opinion, views, and perspectives towards exoskeletons from the associated tweets.

# INTRODUCTION

A robotic exoskeleton is a wearable electromechanical device designed with the main goal of improving the physical performance, stamina, endurance, and capabilities of the person wearing it (Olar et al. 2021). Exoskeleton technology is a novel technology that relies on tracking, measuring, analyzing, and anticipating human behavior and hu-man-machine interactions in multiple ways. Exoskeletons vary in terms of their functionality, mechanism, design, and applications, and there is a relationship between all these parameters that determine the overall working of the exoskeleton (Dollar et al. 2008, Hong et al. 2013). Exoskeletons are classified according to the region they support – for example, upper limb, lower limb, or joint exoskeletons such as the wrist or ankle (Vélez-Guerrero et al. 2021, Li et al. 2021, Castiblanco et al. 2021). Exoskeletons have a variety of uses, including (1) assisted living: assisting the elderly and people with varying forms of disabilities in performing daily routine tasks in an independent manner; (2) military: increasing performance and reducing fatigue; (3) healthcare: improving the wellbeing of individuals who have lost one or both of their arms or legs or who have paralysis or weakness in the same; (4) firefighting: assisting firefighters in climbing rapidly as well as lifting and carrying heavy equipment. Because of its wide range of applications and use cases, exoskeleton technology has advanced rapidly in the last decade and a half. The exoskeleton market was USD 200 million in 2017 (Coren 2021), and it is estimated to become USD 1.3 billion by the end of 2024 (Global Market Insights 2021). With the increasing applications and use cases of exoskeletons, it is crucial to understand user opinion, views, and perspectives towards the same, which would help to study the associated user experience, user acceptance, and user trust. Today's Internet of Everything lifestyle (Langley et al. 2021) is centered around people spending more time on the internet than ever before (Trivedi et al. 2020), and a central aspect of this time is focused on people communicating and sharing their opinion, views, and perspectives towards different topics, usually via social media platforms. Such topics (Huang et al. 2021, Duong et al. 2020) could include emerging technologies, politics, weather, sports, food, travel, games, education, fitness, and health, just to name a few. Twitter, one such social media platform used by people of all age groups, has been rapidly gaining popularity in all parts of the world and is currently the second most visited social media platform (Gruzd et al. 2011). Owing to its constantly increasing popularity, Twitter has been referred to as an online "community" (Bin-Tahir 2021). At present, there are about 192 million daily active users on Twitter (Lin 2021), and approximately 500 million tweets are posted on Twitter every day (Aslam 2021). Therefore, utilizing this immense potential of today's Internet of Everything lifestyle to investigate user opinion, views, and perspectives towards exoskeletons by mining communications from the online "community" of Twitter serves as the main motivation for this work. This paper is organized as follows. Section 2 presents a review of related works in this field. Section 3 presents the methodology and results. Section 4 concludes the paper, which is followed by the author contributions section and references.

## LITERATURE REVIEW

Song et al. (Song et al. 2021) developed a methodology for optimizing the use of an exoskeleton in self-selected walking. In this approach, an algorithm was used to apply torque and estimate the optimal speed pattern for self-selected walking. Zhu et al. (Zhu et al. 2021) developed a triboelectric bi-directional sensor to help simplify and reduce the cost of monitoring motions of users using exoskeletons. This application could detect the rotating, twisting, and bending actions of the user. An index finger exoskeleton was proposed by Sun et al. (Sun et al. 2021) to support the rehabilitation of patients with stroke history. The device was equipped with three motors and utilized a spatial mechanism with a redundant degree of freedom to recognize human-robot axes' alignment. Li et al. (Li et al. 2021) developed a system that employed ordinal and preference feedback, as well as Bayesian posteriors, to estimate the utility landscape of users across four distinct exoskeleton gait parameters. In (Nasiri et al. 2021), the authors developed an adaptive controller that optimized the support provided by an exoskeleton in rehabilitation tasks without prior knowledge of the user's motor capabilities. The approach employed sensory feedback and was applied to the degree of freedom models of the human arm and lower limb to study the overall performance of the system. Hsu et al. (Hsu et al. 2021) proposed a four-degree-of-freedom robotic hip exoskeleton for gait rehabilitation that used active extension and passive abduction to conform with the thigh movements of the user. Sarkisian et al. (Sarkisian et al. 2021) proposed a motorized knee exoskeleton to create a light-weight and compact self-aligning system. The system was tested on 14 healthy participants with the self-aligning mechanism in both locked and unlocked states. The harmony exoskeleton designed by De Oliveira et al. (De Oliveira et al. 2021) could complete shoulder articulation, forearm flexion-extension, and wrist pronation-supination movements. In (Zhou et al. 2021), the authors developed a system that used a multiarticular unpowered exoskeleton to generate positive mechanical work during the stance phase of the user by using the negative mechanical energy of the knee joint. Shi et al. (Shi et al. 2021) proposed a cable-driven 3 degree-of-freedom wrist rehabilitation exoskeleton that was operated by a distributed active and semi-active system. A number of processes and algorithms, such as the rotating compensation mechanism and the assist-as-needed (A.A.N.) algorithm, contributed to the exoskeleton's efficacy in rehabilitation. Gomez-Vargas et al. (Gomez-Vargas 2021) developed the T-FLEX ankle exoskeleton's actuation mechanism, which improved the walking patterns of post-stroke patients. In (Dudley et al. 2021), the researchers devised a technique for determining functional and neuromuscular abnormalities in stroke patients. A damaged hand was used as a subject to test for functional and neuromuscular outcomes with and without a 3D printed passive exoskeleton. Bryan et al. (Bryan et al. 2021) developed an exoskeleton emulator for the hip, knee, and ankle of the user that utilized a mechanism of high torques and powers to assist up-hill runners. The functionality came from the off-board motor used to actuate the entire exoskeleton. The exoskeleton emulator was compatible with a broad range of users and worked efficiently during moving activities. As can be seen from these works, while there has been significant research done in this area in terms of developing, augmenting, and implementing a range of functionalities in exoskeletons, none of these works have focused on investigating user opinion, views, and perspectives towards exoskeletons. Sentiment analysis has shown the potential for interpretation of user opinion, views, and perspectives towards any technology or topic, as seen in prior works (Kaity et al. 2020, Liu et al. 2020, Gorodnichenko et al. 2021) in the field of natural language processing. Therefore, we aim to address this knowledge gap in this field of research by per-forming sentiment analysis of communications on Twitter that involved using sharing their opinion, views, and perspectives towards exoskeletons in the form of tweets. The methodology is discussed in Section 3.

## METHODOLOGY AND RESULTS

A common challenge for working with tweets has been centered around the GET and POST limits of the Twitter API as well as the number of requests that the Twitter API can handle (Rate Limits 2021). The Twitter API can handle a maximum of 900 requests over a 15-minute interval. The limits for the POST statuses/updates are 300 for a 3-hour window, and the GET application/rate_limit_status are 180 per window per user, respectively. Moreover, the Twitter Developer Platform (Standard Search API 2021) has a limit of 7 days on the search index. Or in other words, no tweets can be obtained by using the Twitter Developer Platform for a date that is older than 7 days from the day when the search of tweets is conducted. Collecting tweets related to exoskeletons over a period of 7 days is less likely to provide enough data for performing sentiment analysis of the associated tweets. Therefore, we used one of our prior works (Thakur et al. 2021), which proposes a novel artificial intelligence-based methodology to scrape exoskeleton-based tweets using an intelligent web scraper with Twitter's advanced search API. In this work (Thakur et al. 2021), by using this methodology, we were able to mine about 20,000 tweets over a period of 231 days, from December 1, 2020, to July 19, 2021. The tweets contain diverse forms of communications and conversations related to exoskeletons which communicate user interests, user perspectives, public opinion, reviews, feedback, suggestions, etc., related to exoskeletons. The results were published in the form of an open-access dataset available at (Thakur et al. 2021). Here, we used this dataset to develop a sentiment analysis model using RapidMiner (Mierswa et al. 2006). RapidMiner is a software application development platform that can be utilized to build, execute, and test diverse data science and machine learning algorithms and approaches. The educational version of RapidMiner was used to bypass the data processing limitation of 10,000 rows that is present in the free version of RapidMiner. In RapidMiner, various functions are available in the form of "operators," which can be directly used, customized, developed, or modified to develop an application. A collection of "operators" that have been used to develop an application with one or more output parameters or characteristics is known as a "process". The RapidMiner "process" that was developed for this purpose used the "Extract Sentiment" operator that is available in the "Operator Toolbox" of RapidMiner. This operator performs tokenization of the text (Indurkhya et al. 2010), which is thereafter used to evaluate the sentiment (in terms of positive or negative). In addition to this, the "operator" also assigns a numerical score to the sentence or string by taking into consideration all the tokens. If the numerical value of this score is a non-zero value, then it classifies the string into categories such as fiery, passive, friendly, inspired, etc.

| Row No. | Score | Scoring String | Negativity | Positivity | Uncovered Tokens | Total Tokens | Tweet Text |
|---|---|---|---|---|---|---|---|
| 1 | 0 | | 0 | 0 | 13 | 13 | mashable: This paralyzed ma... |
| 2 | 0 | | 0 | 0 | 11 | 11 | This paralyzed man is using h... |
| 3 | -0.359 | fiery (-0.36) | 0.359 | 0 | 40 | 41 | After yesterday's incident betw... |
| 4 | 0 | | 0 | 0 | 8 | 8 | Replying to @degregori_jack ... |
| 5 | 0 | | 0 | 0 | 7 | 7 | jumping off city buildings with ... |
| 6 | 0.487 | better (0.49) | 0 | 0.487 | 17 | 18 | If you can't handle a powered ... |
| 7 | 0.205 | passive (0.21) | 0 | 0.205 | 15 | 16 | Exoskeleton Market with COVI... |
| 8 | 0 | | 0 | 0 | 6 | 6 | He never once put America first. |
| 9 | 0 | | 0 | 0 | 10 | 10 | Replying to @john_alvz @aw... |
| 10 | 0 | | 0 | 0 | 4 | 4 | Replying to @DT2ComicsChat |
| 11 | 0 | | 0 | 0 | 4 | 4 | Replying to @GardeningAtNigh |
| 12 | 0 | | 0 | 0 | 4 | 4 | Replying to @TristanGinger |
| 13 | 0 | | 0 | 0 | 4 | 4 | Replying to @MacDoesIt |

Figure 1. Results from the RapidMiner "process" that was developed to implement this methodology (the results for all the tweets are not shown here for the paucity of space).

We used the entire dataset of ~20,000 tweets to develop a "process" in RapidMiner that used this "operator," and the results of the same are shown in Figure 1. The result consisted of ~20,000 rows (corresponding to the number of tweets); however, for the paucity of space, only the first 13 rows are represented here. As can be seen from Figure 1, the attribute – "Tweet Text" represents the total text of the tweet. The results that communicate the total tokens, uncovered tokens, positivity, negativity, and the score of the string are represented as different attributes. Figures 2 and 3 show a collection of some of the tweets from this dataset and how the positive and negative sentiments (along with the intensity of these sentiments) associated with these tweets varies in the dataset.

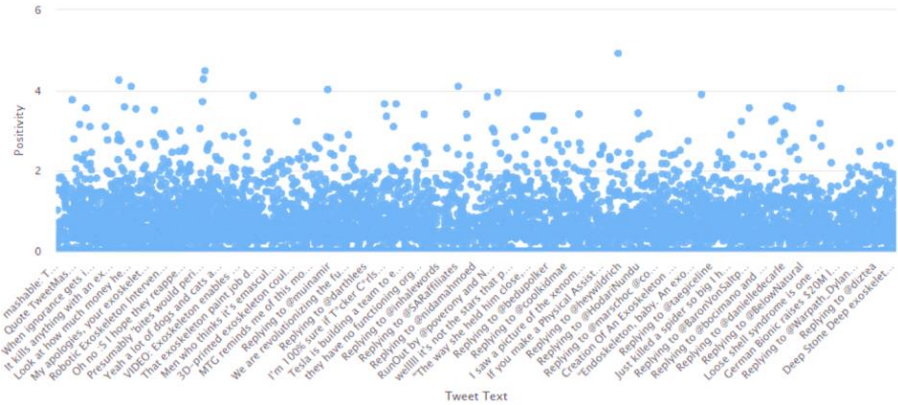

Figure 2. The variation of the intensity of the positive sentiment across different tweets in the dataset (the results for all the tweets are not shown here for the paucity of space).

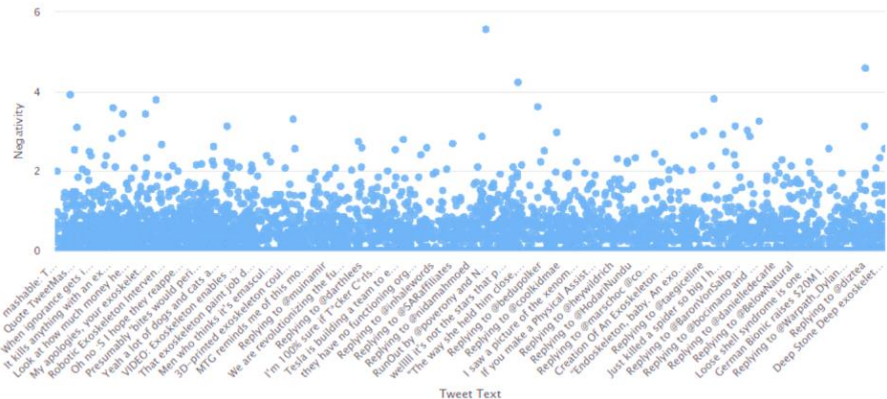

Figure 3. The variation of the intensity of the negative sentiment across different tweets in the dataset (the results for all the tweets are not shown here for the paucity of space).

In Figure 4, we have represented the variation of the numerical values of the overall score of a set of tweets from this dataset. These results demonstrate the efficacy of this approach to analyze and interpret the sentiments associated with tweets related to exoskeletons. The approach also allows calculation of the numerical value of the positive or negative sentiment associated with these respective tweets that allow for interpretation of the intensity of the sentiment. Furthermore, the methodology also assigns a score to the overall tweet by classifying it into categories such as fiery, passive, friendly, inspired, etc.

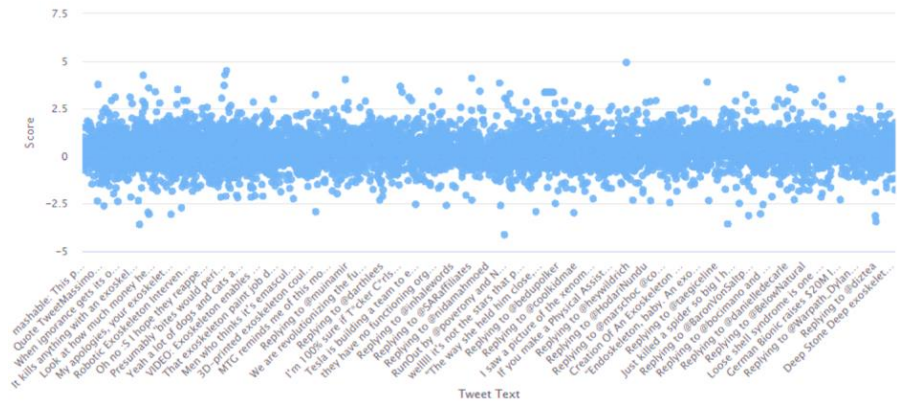

Figure 4. The variation of the overall score of the tweet for different tweets in the dataset (the results for all the tweets are not shown here for the paucity of space).

## CONCLUSION AND FUTURE WORK

The exoskeleton market has been rapidly increasing in the last decade and a half on account of its diverse applications in different domains such as assisted living, military, healthcare, firefighting, and industries. The modern-day exoskeletons have different functionalities to augment human abilities, stamina, potential, and performance, just to name a few, to address different needs in these application domains. In view of this wide-scale applicability and use-cases of exoskeletons, it is crucial to investigate and analyze the public opinion, views, and perspectives towards exoskeletons which would help to interpret the effectiveness of the underlining human-robot, human-machine, and human-technology interactions. The Internet of Every-thing lifestyle of today's living comprises of people spending more time than ever before on the internet sharing their opinion, views, and perspectives, quite often on social media channels. Therefore, capturing such relevant communications related to exoskeletons from social media channels holds the potential for interpretation of user opinions, views, and perspectives towards exoskeletons. To address this challenge, in this work, we have proposed a methodology using the concept of sentiment analysis and natural language processing paradigms. A total of ~20,000 tweets collected over a period of 231 days, from December 1, 2020, to July 19, 2021, were used for testing the efficacy of the proposed approach. It is clear from the results that this method may be used effectively to evaluate and understand the sentiment of tweets about exoskeletons. The method also makes it possible to calculate the numerical value of the positive or negative sentiment linked with the relevant tweets, allowing researchers to investigate the intensity of the sentiment. A score is also assigned to each tweet by this approach, and these scores allow classification of the tweet into categories such as "fiery," "passive," "friendly," and "inspired," among others. Future work would involve developing means to investigate potential reasons related to the negative sentiments associated with certain exoskeleton-based technologies or applications to suggest remedial measures for addressing the same.

## AUTHOR CONTRIBUTIONS

Conceptualization, N.T..; Literature Review, C.L.; Methodology, N.T.; Data Curation, N.T; Formal Analysis, N.T.; Data Visualization and Interpretation, N.T.; Results, N.T.; Writing-Original Draft Preparation, N.T. and C.L; Writing-Review and Editing, N.T; supervision, N.T; project administration, N.T. and C.Y.H.; Funding Acquisition, Not Applicable. All authors have read and agreed to the published version of the manuscript.

## ACKNOWLEDGEMENTS

This research was supported by the University of Cincinnati Graduate School Dean's Dissertation Completion Fellowship.